# First-principles study of the effect of compressive strain on oxygen adsorption in Pd/Ni/Cu-alloy-core@Pd/Ir-alloy-shell catalysts

Jeffrey Roshan De Lile[1], So Young Lee[2], Hyoung-Juhn Kim[2], Chanho Pak[3,*] and Seung Geol Lee[1,*]

[1] *Department of Organic Material Science and Engineering, Pusan National University 2, Busandaehak-ro 63beon-gil, Geumjeong-gu, Busan 46241, Republic of Korea*

[2] *Fuel Cell Research Center, Korea Institute of Science and Technology, Hwarang-ro 14-gil 5, Seongbuk-gu, Seoul 02792, Republic of Korea*

[3] *Graduate Program of Energy Technology, School of Integrated Technology, Institute of Integrated Technology, Gwangju Institute of Science and technology, 123 Cheomdangwagi-ro, Buk-gu, Gwangju 61005, Republic of Korea*

*Corresponding authors:

seunggeol.lee@pusan.ac.kr (S.G. Lee)

chanho.pak@gist.ac.kr (C. Pak)

# Abstract

A palladium-based (Pd-based) core@shell catalyst can be modified to achieve the desired oxygen adsorption properties by selecting an appropriate core composition, surface alloying, and compressive strain. Herein, we present the effects of compressive strain, core composition, and surface alloying in $Pd_3Ni@PdIr(111)$, $Pd_3CuNi@PdIr(111)$, and $Pd_3Cu@PdIr(111)$ alloy-core@alloy-shell catalysts on dioxygen adsorption. Using experimental lattice parameters for the unstrained catalysts, -1% to -5%, strain was systematically introduced. The calculated dioxygen-adsorption energies for the surface Pd and surface Ir atoms reveal that the $Pd_3CuNi@PdIr$ catalyst has the lowest dioxygen-adsorption energy at a given compressive strain. Bader charge calculations show that the $Pd_3CuNi@PdIr$ catalyst surface is the most charge depleted. The d-band model displays an intermediate d-band center downshift for the surface Pd atoms, and the highest downshift for the surface Ir atoms. Due to synergism between charge depletion, the d-band center shift, and the surface alloy effect, the $Pd_3CuNi@PdIr$ catalyst has the lowest dioxygen-adsorption energy. The relationship between the experimentally obtained catalyst-surface mass activity and the theoretically calculated d-band center of the surface Pd and the surface Ir is volcano shaped, with the $Pd_3CuNi@PdIr$ catalyst at the apex of the volcano. The catalytic activities of these catalysts were observed to follow the order: $Pd_3CuNi@PdIr > Pd_3Cu@PdIr > Pd_3Ni@PdIr$. This work sheds light on the importance of ligand and strain effects, as well as surface alloying for the fine-tuning of alloy-core@alloy-shell-catalysts during the rational design of catalysts from first principles.

# 1. Introduction

The world is rapidly embracing new technology to drive the modern economy and curb the negative effects of fossil fuel burning. Fuel cells are a green technology available particularly for automobile applications. Unfortunately, the commercialization of fuel-cell technology is impeded by slow oxygen-reduction kinetics and the high cost of Pt or platinum-group metals (PGM) for use in cathode catalysts. Alternative materials, such as bimetallic and trimetallic transition metals or transition metal-PGM nanoparticles,[1-3] metal-oxides,[4] chalcogenides,[5, 6] transition metal macrocyclic complexes,[7-9] conducting polymers,[10] enzymatic materials,[11] nitrogen-doped carbon,[12] and a variety of carbides[13] have been investigated as Pt replacements. Although these materials show promising oxygen-reduction reaction (ORR) activities, their stabilities under harsh fuel-cell operating conditions are less than expected.

The contemporary development of core@shell catalysts with random alloys in their core structures has paved the way to more-stable alternative catalysts. The electronic properties of these materials are highly sensitive to the core alloy composition; therefore, these catalytic materials can be fine-tuned by varying their core-component compositions (i.e., the ligand effect).[14] This effect is a consequence of electronic charge redistributions between core and shell atoms due to dissimilar metal atoms present in the core and shell. Zhang et al.[15] demonstrated this effect by varying the core composition from pure Pd to 20%, 40%, 60%, 80%, and 100% Cu with a Pd shell. This approach reduced the magnitude of the oxygen-adsorption energy, with its value changing from -1.7 eV (Pd@Pd) to -1.2 eV (Cu@Pd). A recent publication by Ham et al.[16] reported the use of ligands to segregate the surface from subsurface Co atoms; these authors introduced an iridium (Ir) layer to prevent Co migration to the surface, which improved the stability of the ternary alloy. Averting the undesirable dissolution of active metals on the surface is particularly important. The introduction of compressive strain reduces the bonding distances between shell atoms and increases the overlap between d-states, which downshifts the d-band center with respect to the Fermi level. Compressive strain reduces the ability of the catalyst surface to adsorb oxygen and improves the overall kinetics of the ORR. A recent publication[17] revealed how lanthanides can be used to contract the Pt-Pt distance and induce compressive strain in a catalyst in order to improve the ORR. The relationship between activity and lattice parameter was volcano-shaped with $Pt_5Tb$ (platinum-terbium) and $Pt_5Gd$ (platinum-gadolinium) at the apex of the volcano. The authors claimed that they enhanced activity by a factor of three-to-six through the introduction of compressive strain. Therefore, the ligand effect and the strain effect have often been used to tune the catalytic properties of core@shell catalysts. Nevertheless, an alloying shell may also change the electronic properties of the surface atoms and affect the ORR. When Pt was alloyed with various transition metals, such as Cu,[18-20] Ni,[21-23] Co,[24, 25] and Fe,[21, 23, 24] lower oxygen-binding energies and enhanced ORR activities were observed.[26] A similar study reported a large database of stable oxygen-reduction catalysts for Pd including, $Pd_2PdSb$, $Pd_2PdAg$, $Pd_2PdAs$, $Cu_2PdCu$, $Pd_3Ru$, and $Pd_3Re$.[27] Recently, a novel PdCu(Ni)@PdIr alloy-core@alloy-shell catalyst for a high-temperature fuel cell was reported.[28] Experimental catalyst systems with PdNi, PdCuNi, and PdCu alloys in their core compositions combined with Pd-Ir alloy shells have been synthesized and tested under fuel-cell operating conditions. The authors claim that the PdCuNi@PdIr catalyst performed best towards the ORR; however, the theoretical foundation of performance is not well understood. Some theoretical work claims that core@shell catalysts fully relax their surface shell structures; consequently, compressive or tensile strain does not play a significant role in tuning catalytic properties.[29] On the other hand, some literature claims that the ligand effect is more local and vanishes quickly after two or three atomic layers; hence the strain effect is more generally useful when designing catalysts.[30, 31] Moreover, the Cu and Ni ions in the core structures have similar properties; hence, the ligand effect is expected to be similar in these catalysts.

In this study, we investigated the effect of compressive strain, the ligand, and surface alloying on oxygen adsorption by exploring the (111) surfaces of Pd$_3$Ni@PdIr, Pd$_3$CuNi@PdIr, and Pd$_3$Cu@PdIr catalysts with experimentally obtained lattice constants using density functional theory (DFT). The oxygen-binding trends and the d-band center shifts with respect to the Fermi level were systematically examined, with compressive strains that ranged from 0% to -5%. Correlations between compressive strain, the electronic structure, and the oxygen-adsorption properties were determined. We observed that the ligand effect dictates oxygen adsorption; however, surface alloying introduces competition between the surface Pd and Ir atoms for oxygen adsorption, with the Ir tending to adsorb oxygen more strongly. Hence, we surmise that the inferior catalytic performance of these alloy-core@alloy-shell catalysts compared to the commercial PtCo catalyst is due to the higher dioxygen-adsorption energies at the surface Ir atoms. Finally, we constructed a volcano plot to determine the most-active catalyst for the ORR using the Brønsted–Evans–Polanyi linear scaling relationship[32] that exists between mass activity and the d-band center. We propose a method for improving the Pd-dominated catalytic activity by selecting suitable elements that maximize the electronegativity difference between Pd and the alloy surface metal whereby the oxygen-adsorption energy on the alloy metal is reduced.

## 2. Computational details

Based on our previous work, it was demonstrated that for bulk composition, the atomic ratios of Pd, Ir, Ni, Cu were 1:0.154:0.183:0 for PdNi@PdIr catalysts, 1:0.159:0.101:0.121 for PdCuNi@PdIr catalyst, and 1:0.109:0:0.198 for PdCu@PdIr catalyst, respectively[28]. That is in the bulk 6.5 Pd atoms to one Ir atom, and five Pd atoms to one Cu and one Ni atom in approximate composition. Nevertheless, Pd presents in both core and shell layers. Thus, we took the liberty as this is a theoretical work to split three Pd atoms into the core and three atoms into the shell. Therefore, we have chosen $Pd_3Ni$, $Pd_3CuNi$, and $Pd_3Cu$ as core compositions to represent higher amount Pd. Similarly, in shell structure, $Pd_3Ir$ alloy composition was chosen to represent the higher amount of Pd in the shell. However, lack of explicit crystallographic information about the structures of PdX(Y) catalysts, we assumed the face-centered-cubic (FCC) $L1_2$ ordered alloy structure for the $Pd_3Ni$, $Pd_3CuNi$, and $Pd_3Cu$ core structures of the alloy-core@alloy-shell catalysts with base metals (Cu, Ni) at the corners and the noble metal (Pd) centered on the faces.[33] The supercell, random structure generating software was used to test the available structures.[34] This approximation created stable structures for $Pd_3Ni$, $Pd_3CuNi$, and $Pd_3Cu$. The Pd-Ir monolayer ordered alloy surface was then created on top of each core structure to generate the alloy-core@alloy-shell catalysts. The shell layer thickness also alters the oxygen adsorption energy in catalysts.[35, 36] In our previous experimental paper[28], we attempted to understand the core composition and strain effect on catalyst performance without altering the shell layer thickness. Therefore, in this theoretical work, we didn't attempt to change the shell thickness, hence monolayer Pd-Ir shell was chosen. Each structure consisted of twelve atoms with eight core atoms and four surface atoms. There were three layers in each alloy-core@alloy-shell structure and the bottom core layer in one set of structures were fixed at the experimental lattice parameters (3.851 Å, 3.844 Å, and 3.839 Å for $Pd_3Ni$@PdIr, $Pd_3CuNi$@PdIr and $Pd_3Cu$@PdIr respectively), while the lattice parameters of the bottom core layers of the other structures were fixed at compressive strains of -1%, -2%, -3%, -4%, and -5% with respect to the experimental values.

In our surface strain model, the primary assumption is that the compressive strain on the surface is similar to the strain on the core atoms. In the experimental core-shell catalysts, the shell layer adopted smaller lattice parameter of the core atoms.[28] As a result, the shell layer was compressively strained. To do that theoretically, we constrained the bottom core layer to fix at lattice parameter to represent -1%, -2%,-3%,-4% and -5% compressive strains with respect to the unstrained catalysts. However, after optimization, the surfaces were slightly expanded due to surface relaxation. If we calculate the bond length variation of the surface Pd atoms of Pd$_3$CuNi@PdIr catalyst from 0% to -1% strain using the measured bond lengths after surface relaxation (see Table-1), this surface experienced -0.88% strain, which was slightly smaller than -1%. Similarly, at higher strain values actual surface strains were -1.77%, -2.69%, -3.60%, -4.55%, respectively. If we add more surface layers, these relaxations become substantially larger and our assumption is no longer valid. Moreover, the ligand effect only extended to a very thin shell, particularly to a single atomic layer.[37-39] Thus, to satisfy the above-mentioned conditions, the monolayer Pd-Ir shell was selected.

The supercells were constructed with 15 Å vacuum layers in the Z-direction to avoid interactions with periodic self-images. The calculations were executed using the Vienna *ab initio* Simulation Package (VASP),[40] at the GGA level of theory using the Perdew-Burke-Ernzerhof (PBE) functional.[41, 42] To determine equilibrium geometries, the ground state of each structure was located using spin-unrestricted calculations, with the forces reduced to less than 0.005 eV/Å. Although we studied catalysts systems of relevance to high-temperature fuel cells, their operating temperatures were varied between 150 and 200 °C, which are lower than the Curie temperature of Ni (357.85 °C); this approach is justified as the spin-induction effect still prevails at these operating temperatures.[43] Van der Waals forces are important for the adsorptions of small molecules onto metal surfaces, therefore the DFT-D3 correction was applied to all calculations. The projector-augmented wave method was used to describe the core electrons, and the plane-wave basis set cutoff energy was set to 400 eV to represent valence electrons. A 12 × 12 × 1 Monkhorst-Pack grid[44] was chosen for Brillouin zone sampling in density-of-state calculations in order to determine the d-band center, whereas a 5 × 5 × 1 Monkhorst-Pack grid was used during initial geometry optimizations. The

VTSTTOOLS method developed by Henkelman et al.[45] was used to calculate d-band centers. The -7 to 5 eV energy range was chosen for predicting the d-band centers of all catalyst structures. Dioxygen adsorption was examined at the surface-Pd top site, hollow sites, and on the top site of the surface Ir. Dioxygen was observed to adsorb end-on, and the dioxygen-adsorption energy was calculated using the following equation:

$$E_{ads} = E_{O2-Pd3X(Y)} - (E_{Pd3X(Y)} + E_{O2}) \tag{1}$$

where $E_{O2-Pd3X(Y)}$ is the gas-phase total energy of the alloy-core@alloy-shell catalyst with adsorbed dioxygen, $E_{Pd3X(Y)}$ and $E_{O2}$ are the gas-phase energies of the bare alloy-core@alloy-shell catalyst surface and dioxygen in its triplet ground state, respectively. Based on this equation, a more negative energy value indicates stronger adsorption. The catalyst structures used in this study are displayed in Figure 1. The end-on binding of the oxygen molecule is observed on all the catalysts surfaces, as shown in Figure 2, in which Figures 2(a) and (b) display dioxygen binding to the surface Pd and Ir atoms of the Pd$_3$Cu@PdIr catalyst, and Figures 2(c) and (d), and 2(e) and (f) show dioxygen binding at the surface Pd and surface Ir of the Pd$_3$CuNi@PdIr and Pd$_3$Ni@PdIr catalysts, respectively. The relaxed structures are used to determine the strains and dioxygen adsorption energies.

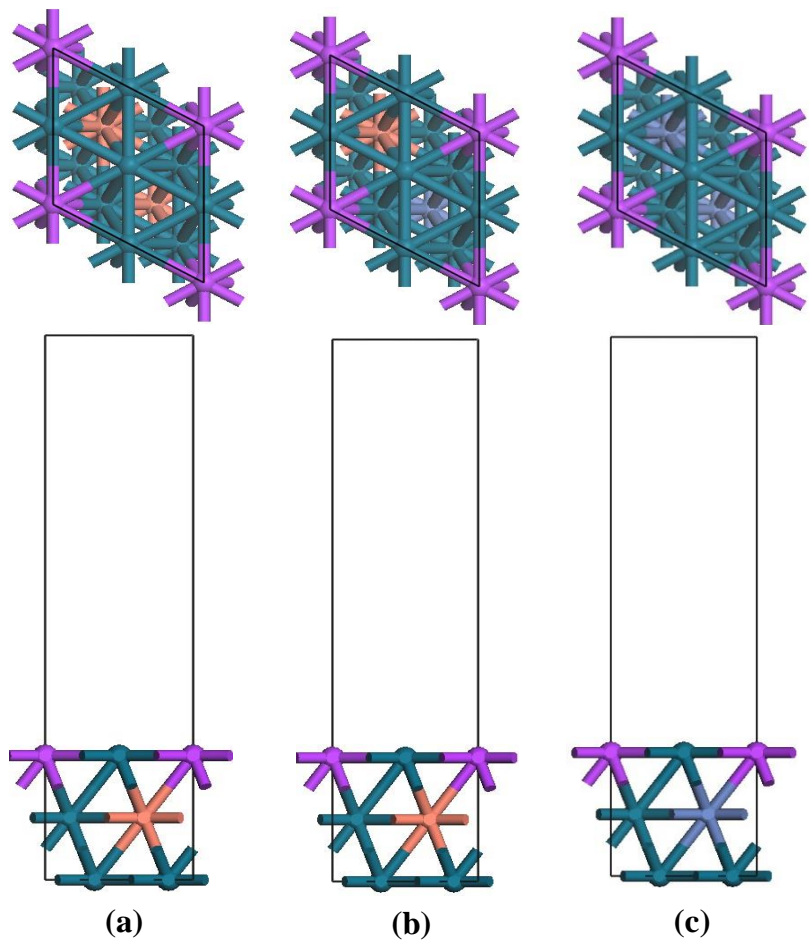

**Figure 1**. Core@shell surface models: (2×2) structures of (a) $Pd_3Cu$@PdIr, (b) $Pd_3CuNi$@PdIr and (c) $Pd_3Ni$@PdIr. Color scheme: purple, Ir; turquoise, Pd; orange, Cu; and light blue, Ni. The 15 Å vacuum layer and simulation box are represented as black lines in the bottom images, while the surfaces are presented in the upper images.

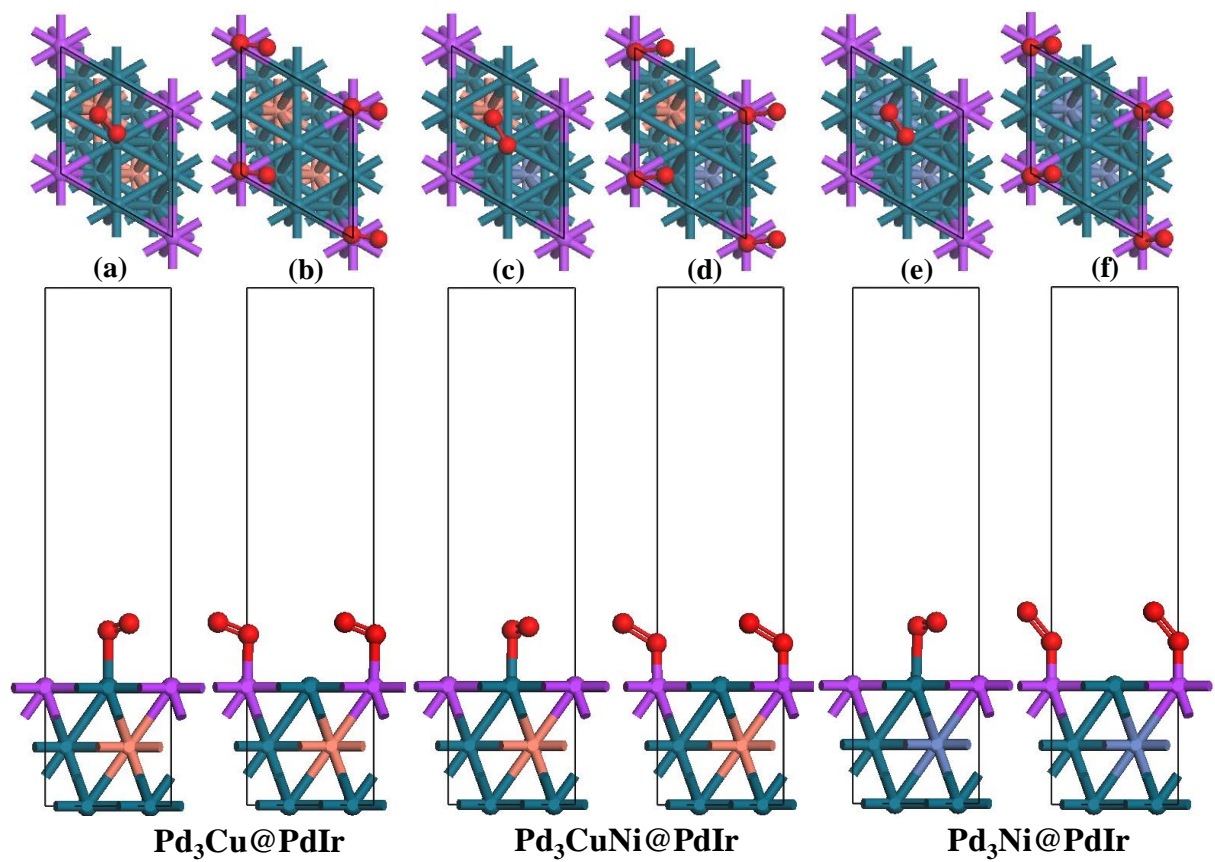

**Figure 2**. Dioxygen binding to the catalyst surfaces. (a) and (b) Dioxygen binding to the surface Pd and the surface Ir of the $Pd_3Cu@PdIr$ catalyst. (c)-(d) Dioxygen adsorption by the $Pd_3CuNi@PdIr$ catalyst and the (e)-(f) $Pd_3Ni@PdIr$ catalyst. Each catalysts surface adsorbs molecular oxygen in an end-on fashion. Top views of the surfaces are displayed in the top images, while side views are displayed below, with the simulation cell highlighted in black. Note that the four oxygen molecules bound to the Ir atoms are a result of the periodic boundary condition. Color scheme: purple, Ir; turquoise, Pd; orange, Cu; light blue, Ni; and red, oxygen.

## 3. Results and Discussion

*3.1 Dioxygen-adsorption energies and d-band center analyses*

The relationship between the composition of the catalyst core and the dioxygen-adsorption energy at the surface Pd at different strain values shows demonstrably lower adsorption energies for the $Pd_3CuNi@PdIr$ catalyst, as shown in Figure 3(a). The $Pd_3Cu@PdIr$ catalyst heavily competes with the $Pd_3CuNi@PdIr$ catalyst at higher compressive strain values;

consequently, the dioxygen-adsorption energy of the Pd$_3$Cu@PdIr catalyst rapidly approaches that of the Pd$_3$CuNi@PdIr catalyst. However, the Pd$_3$Ni@PdIr catalyst lags behind the other two catalysts at each point, except at the highest (-5%) compressive strain, where it exhibits a higher dioxygen-adsorption capacity. Therefore, we cannot expect to see a significant difference in the dioxygen-adsorption energies at the surface Pd atoms in these catalysts systems at compressive strains above -5%. Figure 3(b) shows the dioxygen-adsorption energy at the surface Ir atom as a function of core composition, which exhibits a similar trend to that observed for the surface Pd, however, the Pd$_3$Ni@PdIr catalyst shows the highest adsorption energies at all compressive strain values. In addition, the Pd$_3$CuNi@PdIr and the Pd$_3$Cu@PdIr catalysts show lower adsorption energies.

The d-band center at different core compositions is dependent on compressive strain, as shown in Figures 3(c) and (d). The highest d-band center downshift with respect to the Fermi level for the surface Pd atom was observed for the Pd$_3$Cu@PdIr catalyst, followed by the Pd$_3$CuNi@PdIr catalyst and the Pd$_3$Ni@PdIr catalyst (see Figure 3(c)). In contrast, the surface Ir atom exhibits the highest downshift in the Pd$_3$CuNi@PdIr catalyst, with the lowest downshift (the highest upshift of the d-band center) observed for the Pd$_3$Ni@PdIr catalyst (see Figure 3(d)); the downshift of the Pd$_3$Cu@PdIr catalyst lies between those of these two catalysts. This d-band trend directly correlates with the energy trend for dioxygen adsorption at the surface Ir. Hence, the catalyst with the highest d-band downshift is fairly straightforwardly associated with the lowest dioxygen adsorption for the surface Ir atoms. However, the dioxygen adsorption trend does not directly correlate with the observed d-band downshifts for the surface Pd atoms in these catalysts.

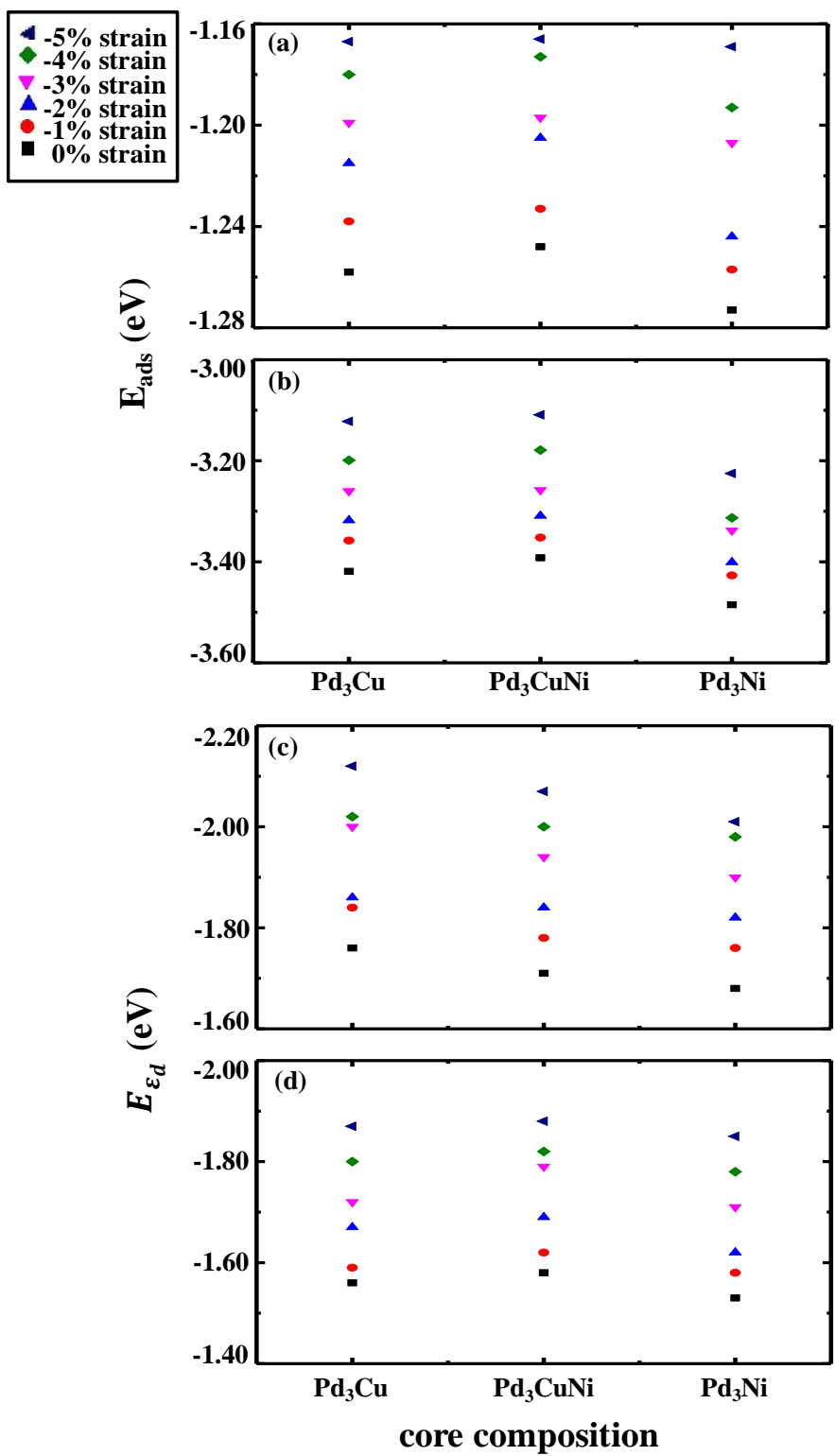

**Figure 3.** Energies for the adsorption of dioxygen at: (a) the surface Pd and (b) surface Ir. The d-band centers of: (c) the surface Pd and (d) the surface Ir, as functions of the catalyst core composition. The legend shows the applied compressive strains of the experimentally found lattice-parameter values.

In order to understand the effect of compressive strain on these d-band center shifts, we determined the surface Pd to Ir and Pd to Pd bond distances, the results of which are listed in Table 1. The surface Pd to Pd distance systematically decreases with increasing compressive strain. Moreover, due to the Pd$_3$Cu@PdIr catalyst having the smallest lattice parameter value (see Table 1), it exhibits the shortest bond distance among all three catalysts at a given compressive strain, which is ascribable to the smaller atomic radius of the Cu atom (145 pm) compared to that of Ni (149 pm)[46]. The Pd$_3$CuNi@PdIr catalyst exhibits intermediate bond distances, while the longest surface Pd-Pd bond distances are observed for the Pd$_3$Ni@PdIr catalyst. Clearly, shorter bond distances enhance overlap between the surface Pd-Pd d-orbitals the most and downshift the d-band significantly, whereas an intermediate Pd-Pd bond distance is associated with an intermediate downshift of the respective d-band center, and the longest bond lengths result in the highest upshift of the d-band center with respect to the Fermi level. As a result, the Pd$_3$Cu@PdIr catalyst exhibits the highest d-band center downshift, while the Pd$_3$Ni@PdIr catalyst exhibits the highest d-band center upshift, and that of the Pd$_3$CuNi@PdIr catalyst lies between the two. Hence, the trend predicted on the basis of the surface Pd-Pd bond distance is perfectly matched with the calculated d-band-center trend for the surface Pd atoms. A similar trend was observed for the surface Pd-Ir bond distances. An electronic structure origin for the observed changes in the d-band center is not visible for individual catalyst systems in the surface-Pd and -Ir density-of-state (DOS) diagrams shown in Figure 4. However, the DOS diagrams do reveal downshifts in the d-band center at 0%, -2% and -4% compressive strains, as depicted in Figures 4(a)-(c) for the surface Pd, and 4(d)-(f) for the surface Ir atoms. In addition, the surface Ir atom shows significantly higher DOS states at or close to the Fermi level; hence, Ir can transfer charge more effectively to molecular oxygen, which explains the higher energy associated with dioxygen adsorption at the surface Ir atoms. For the surface Pd, the d-band center downshift is more prominent and it ranges from -1.72 eV (0% compressive strain) to -2.07 eV (-5%) on average. This range is -1.56 eV (0%) to -1.87 eV (-5%) on average for the surface Ir (see Figures 3(c) and (d)). Thus, the higher d-band center upshift provided additional information to support the higher dioxygen adsorption energy on the surface Ir. A similar trend was reported in the literature, which gives credence to this work.[47]

**Table 1.** Catalysts and their calculated properties.

| Catalyst | Strain | Lattice constant/Å | Pd-Pd/Å | Pd-Ir/Å | Charge transfer* Pd/Ir | $O_{ads}$/eV Pd | $O_{ads}$/eV Ir |
|---|---|---|---|---|---|---|---|
| Pd@PdIr** | | 3.891 | 2.748 | 2.758 | -0.042/+0.363 | -3.53 | -3.58 |
| Pd$_3$Cu@PdIr | 0 % | 3.839 | 2.703 | 2.719 | +0.054/+0.237 | -1.26 | -3.42 |
| Pd$_3$CuNi@PdIr | | 3.844 | 2.705 | 2.722 | +0.051/+0.099 | -1.25 | -3.39 |
| Pd$_3$Ni@PdIr | | 3.851 | 2.708 | 2.726 | +0.073/+0.299 | -1.27 | -3.49 |
| Pd$_3$Cu@PdIr | -1% | 3.801 | 2.680 | 2.692 | +0.046/+0.101 | -1.24 | -3.36 |
| Pd$_3$CuNi@PdIr | | 3.806 | 2.681 | 2.696 | +0.042/+0.059 | -1.23 | -3.35 |
| Pd$_3$Ni@PdIr | | 3.812 | 2.684 | 2.699 | +0.064/+0.135 | -1.26 | -3.43 |
| Pd$_3$Cu@PdIr | -2% | 3.763 | 2.656 | 2.666 | +0.037/+0.074 | -1.22 | -3.32 |
| Pd$_3$CuNi@PdIr | | 3.767 | 2.657 | 2.669 | +0.014/+0.043 | -1.21 | -3.30 |
| Pd$_3$Ni@PdIr | | 3.775 | 2.659 | 2.672 | +0.057/+0.093 | -1.24 | -3.40 |
| Pd$_3$Cu@PdIr | -3% | 3.724 | 2.631 | 2.640 | +0.022/+0.064 | -1.20 | -3.26 |
| Pd$_3$CuNi@PdIr | | 3.729 | 2.632 | 2.643 | +0.007/+0.036 | -1.19 | -3.25 |
| Pd$_3$Ni@PdIr | | 3.736 | 2.634 | 2.646 | +0.048/+0.082 | -1.21 | -3.34 |
| Pd$_3$Cu@PdIr | -4% | 3.686 | 2.605 | 2.615 | +0.017/+0.045 | -1.18 | -3.20 |
| Pd$_3$CuNi@PdIr | | 3.690 | 2.607 | 2.618 | +0.001/+0.032 | -1.17 | -3.18 |
| Pd$_3$Ni@PdIr | | 3.697 | 2.609 | 2.620 | +0.034/+0.080 | -1.19 | -3.31 |
| Pd$_3$Cu@PdIr | -5% | 3.647 | 2.578 | 2.591 | +0.006/+0.034 | -1.17 | -3.12 |
| Pd$_3$CuNi@PdIr | | 3.652 | 2.582 | 2.593 | +0.001/+0.021 | -1.16 | -3.10 |
| Pd$_3$Ni@PdI | | 3.659 | 2.584 | 2.595 | +0.009/+0.046 | -1.18 | -3.22 |

*Charge-transfer values obtained from Bader charge analyses are also included for the surface Pd and Ir atoms (Pd/Ir); a negative value indicates charge accumulation on the respective atom, while a positive value indicates charge depletion.

** For comparison, data for the catalyst with a pure Pd core and a Pd-Ir shell (Pd@PdIr) are also listed.

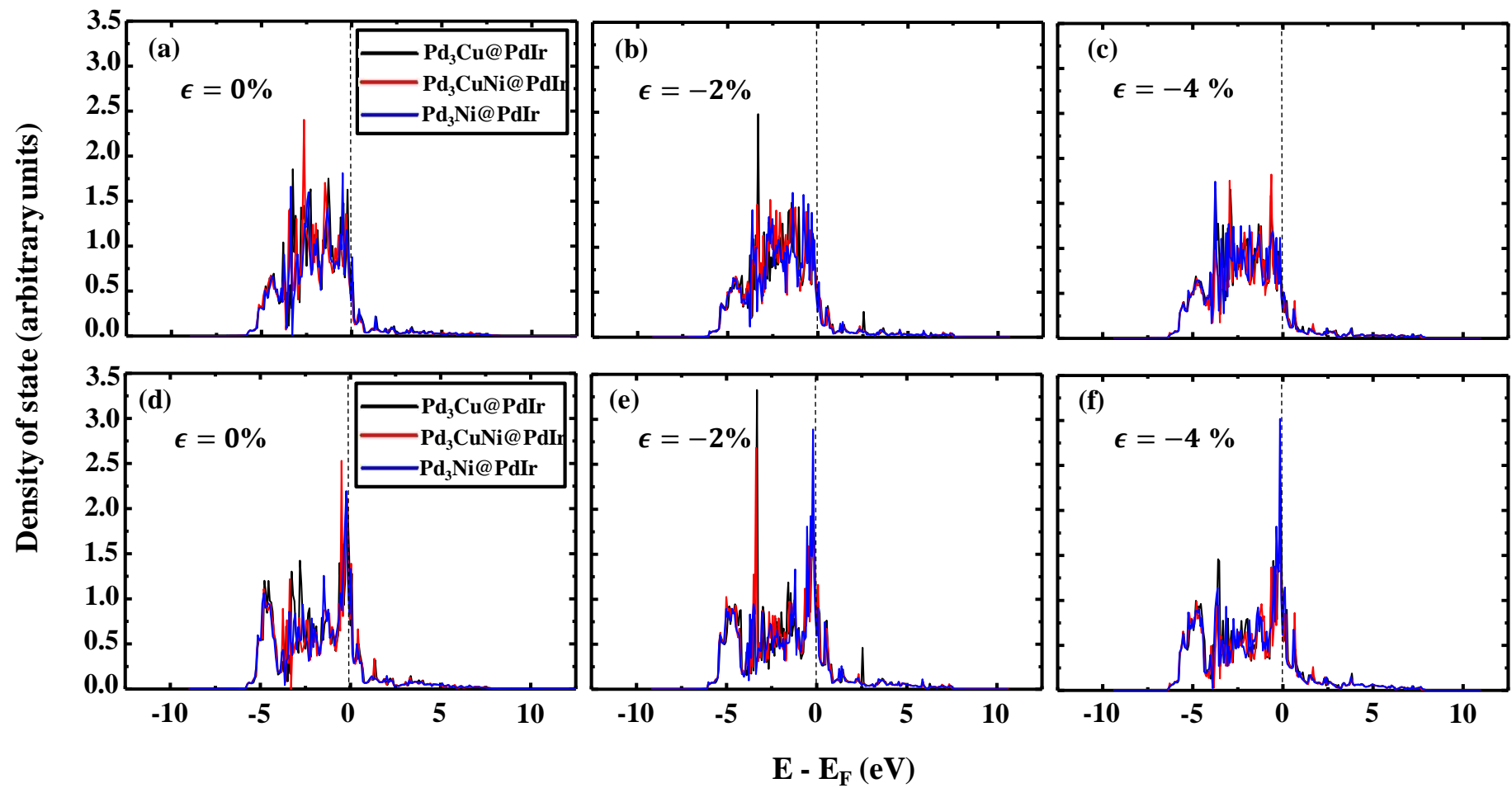

**Figure 4.** Densities of state (DOS) of: (a)–(c) the surface Pd and (d)–(f) the surface Ir atoms with 0%, -2%, and -4% strain. We chose the DOS of the dioxygen-adsorbed surface atom to be representative.

*3.2 Dioxygen-adsorption energies and charge analyses*

Figure 5 shows relationships between compressive strains and calculated dioxygen-adsorption energies at the surface Pd and the Ir atoms. Dioxygen-adsorption energies for the surface Pd and the Ir atoms of a pure Pd core with a Pd-Ir surface (shell) are also included in Figure 5(b), with the calculated dioxygen-adsorption energies presented in Table 1. The Figure 5(a) illustrates all the dioxygen binding energies on the surface Pd atoms converged to (almost) a similar value at the -5% compressive strain. Consequently, further compression would deviate the dioxygen binding energies of the catalyst systems from the current trend. In addition, theoretical work related to compressive strain controlled catalyst activity often restricted to -3% or -5% compressive strains.[48-50] As a result, 0% to -5% compressive strain range is suitable to describe our model catalysts systems. The surface Pd atoms exhibit dramatically lower dioxygen-adsorption energy magnitudes when the pure Pd core is replaced with a Cu and/or Ni alloy, with values of -3.53 eV to -1.26 eV on average for the three unstrained catalysts in this study. However, the surface Ir atoms do not show such a significant change, namely -3.58 eV for the unstrained catalyst with the pure Pd core, and an average of -3.43 eV after alloying. Therefore, on average, the magnitude of the dioxygen-adsorption energy at the surface Pd is reduced by 2.27 eV due to alloying, while the analogous energy for the surface Ir is reduced by 0.15 eV. On the other hand, as the compressive strain of the alloyed core is increased from -1% to -5%, the magnitude of the oxygen-adsorption energy at the surface Pd is reduced by 0.09 eV, while the analogous decrease for the surface Ir is 0.28 eV. Therefore, the significantly weaker dioxygen adsorptions at the surfaces of the alloyed core catalysts can be ascribed to the ligand effect resulting from the composition of the core. Using the Bader charge-analysis scripts from the Henkelman group,[45] we calculated the charge redistributions at the surface Pd and surface Ir atoms in the Pd@PdIr, Pd$_3$Ni@PdIr, Pd$_3$CuNi@PdIr, and Pd$_3$Cu@PdIr catalysts in order to understand the ligand effect, the results of which are presented in Table 1. Excess electrons accumulate at the surface Pd atoms when the core is composed of pure Pd atoms; hence, the Pd atoms are more negatively charged and more easily transfer this excess charge to molecular oxygen. Consequently, the dioxygen-adsorption energy is higher. However, alloying with Cu and/or Ni results in the loss of the excess charge on the surface Pd, which becomes positively charged (charge depletion). The highest charge depletion at the surface Pd was observed for the $Pd_3$CuNi@PdIr catalyst; therefore, it shows the weakest dioxygen adsorption, despite its intermediate d-band center downshift. In addition, charge depletion at the surface Pd is more severe than that at the surface Ir, which is in agreement with the calculated DOS results, since the surface Ir atom possesses higher state densities closer to the Fermi level that facilitate charge transfer. Therefore, surface Ir atoms bind more strongly to dioxygen than surface Pd atoms. Moreover, based on the calculated Bader charge-transfer values, the lowest positive charges were found at the surface Ir atoms of the $Pd_3$CuNi@PdIr catalyst. This charge depletion also results in the weakest dioxygen adsorption at the surface Ir atoms of the $Pd_3$CuNi@PdIr catalyst. Furthermore, higher compressive strain leads to lower charge transfer at the surface atoms. The relationship between the d-band center downshift and the dioxygen-adsorption energy is linear throughout the entire compressive-strain range for these catalysts, as revealed in Figures S1(a) and (b). Figure S1(a) shows the relationship between the d-band center and the dioxygen-adsorption energy at the surface Pd, while Figure S1(b) shows the analogous relationship for the surface Ir. As a general trend, higher d-band center downshifts lead to lower dioxygen-adsorption energies at the catalyst surfaces. This result is in good agreement with literature-reported core@shell catalyst trends.[30, 49] The lower amount of charge transfer and the downshift of the d-band center at higher strains lower the magnitude of the dioxygen-adsorption energy. Clearly, a subtle interplay between the ligand effect and the strain effect controls dioxygen binding on these catalysts.

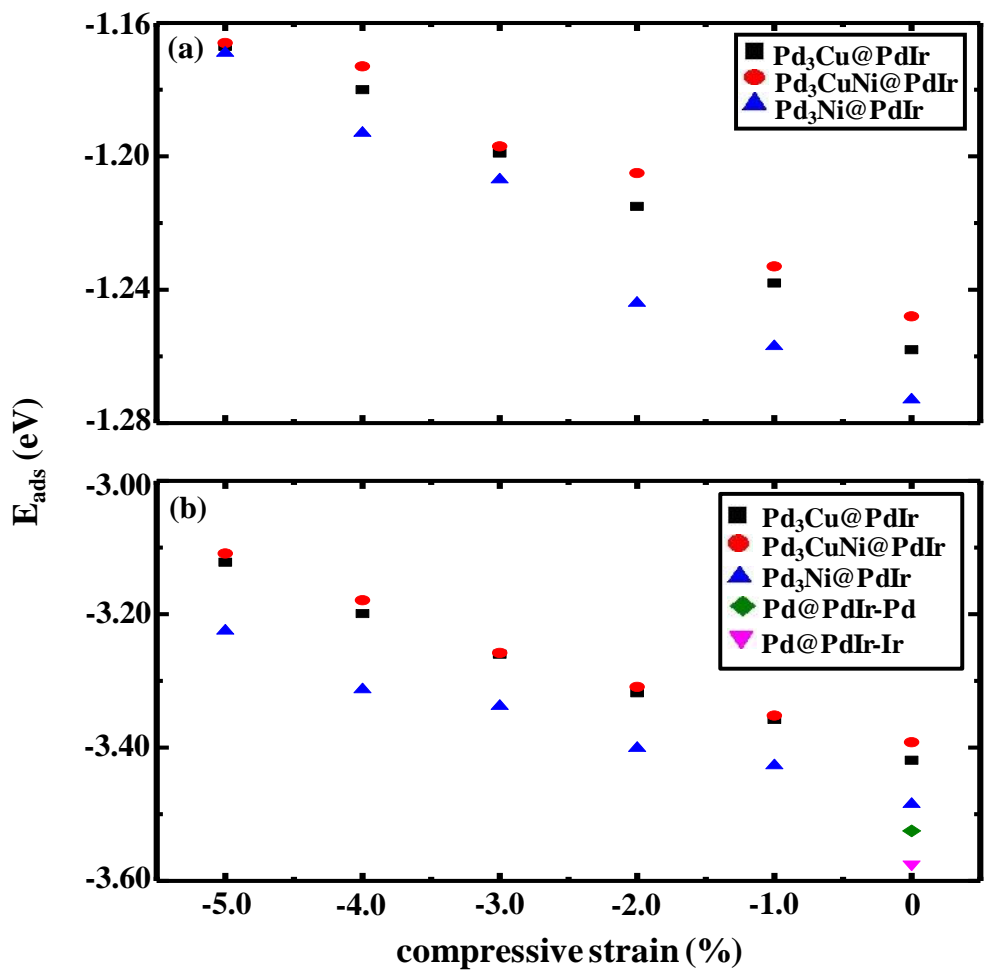

**Figure 5.** Dioxygen-adsorption energies as functions of compressive strain at: (a) the surface Pd and (b) the surface Ir atoms. The surface Pd tends to bind dioxygen weakly, while surface Ir exhibits very strong binding. Panel (b) also shows dioxygen binding for a pure Pd core, where the green symbol indicates dioxygen binding at the surface Pd atoms, and pink symbol indicates dioxygen binding at the surface Ir atoms.

*3.3 Effect of the alloyed Ir on the catalyst surface*

Although the electronegativity of both the Ir and the Pd metals are 2.20 on the Pauling scale,[46,][51] the surface Ir atoms strongly adsorb dioxygen, as is clearly shown by our oxygen-adsorption energy calculations. Therefore, oxygen adsorption is dominated by the surface Ir atoms rather than the surface Pd atoms in these catalytic systems. As a consequence, compellingly lower dioxygen adsorption is achieved at the surface Pd, which impairs the overall performance of the catalyst. Hence, the lower ORR activity of the Pd$_3$CuNi@PdIr catalyst compared to that of the PtCo TKK catalyst is attributable to the surface Ir that dominates dioxygen adsorption. The DFT calculations of the Pd(111) and PdIr(111) surfaces with the Pd$_3$CuNi core at 0% compressive strain revealed further surface alloying effects with the Ir. The dioxygen adsorption energies on those two different shells are -0.85 eV for the Pd$_3$CuNi@Pd(111) and -1.25 eV for the Pd$_3$CuNi@PdIr(111). Moreover, the ground state energies of these catalysts are -62.26 eV and -65.24 eV for the Pd$_3$CuNi@Pd(111) and Pd$_3$CuNi@PdIr(111), respectively. Therefore, the surfae alloying of the Ir improves the stability as well as dioxygen adsorption energy of the Pd$_3$CuNi@PdIr(111) catalyst. In addition, those ground state energies indicate that the Ir is segregated to the topmost catalyst surface layer.

The Pt$_3$Y, Pt$_5$Y, Pt$_3$Hf, Pt$_3$Sc, Pt$_3$Zr, and Pt$_5$La catalysts[52] have recently been reported to perform well in the oxygen-reduction reaction; among them Pt$_3$Y, Pt$_5$Y, and Pt$_5$La are particularly effective.[52] The electronegativity of these metals on the Pauling scale are: 1.1 (La), 1.2 (Y), 1.3 (Hf), 1.33 (Zr), 1.36 (Sc), and 2.23 (Pt),[46, 51] which reveals that Pt$_3$Y, Pt$_5$Y, and Pt$_5$La have the highest electronegativity differences between their constituent metals among these catalysts. We believe that the preferred alloyed metals have lower dioxygen-adsorption energies, which enhance the ORR. Therefore, we propose that alloying Pd with less-electronegative metals on the catalyst surface, such as Y or La, leads to better performance.

The alloyed core elements, namely Cu and Ni, are segregated from the topmost surface layer;[53, 54] steric factors in these catalysts prevent these atoms from moving to the surface due to the large size of Ir (180 pm atomic radius) compared to the surface Pd atoms (169 pm).[46] Hence, these proposed metals may effectively segregate the core atoms due to their large atomic radii. Furthermore, we observed a volcano-shaped relationship between the mass activity of the catalyst and the lattice parameter, as shown in Figure S2. The theoretically calculated d-band center variations are directly related to the value of the lattice parameter, as the lattice constant dictates the surface bond distances. Therefore, taking the experimentally determined mass activities of the catalyst surfaces (Pd + Ir) and the theoretically calculated combined d-band centers for both Pd and Ir into account, we constructed a volcano plot of catalytic activity, as shown in Figure 6. The $Pd_3CuNi@PdIr$ catalyst lies at the apex of the plot, and the $Pd_3Cu@PdIr$ and the $Pd_3Ni@PdIr$ catalysts reside on the left and right sides of the $Pd_3CuNi@PdIr$ catalyst, respectively. Hence, through synergism between the ligand effect, the d-band center shift, and the surface alloying effect, the $Pd_3CuNi@PdIr$ catalyst exhibits the poorest dioxygen adsorption and, consequently, the best catalytic ORR performance. The $Pd_3Cu@PdIr$ catalyst surpasses the vortex of the volcano, hence its catalytic activity is compromised due to high dissociation barriers.[15, 32] On the contrary, the $Pd_3Ni@PdIr$ catalyst has lower activity due to strong dioxygen adsorption, which may ultimately lead to product-removal difficulties and a lower catalyst reaction rate.[15, 32] Therefore, the theoretically predicted catalytic-activity trend is in good agreement with the experimentally discovered trend; i.e., $Pd_3CuNi@PdIr$ > $Pd_3Cu@PdIr$ > $Pd_3Ni@PdIr$, as reported previously.[28]

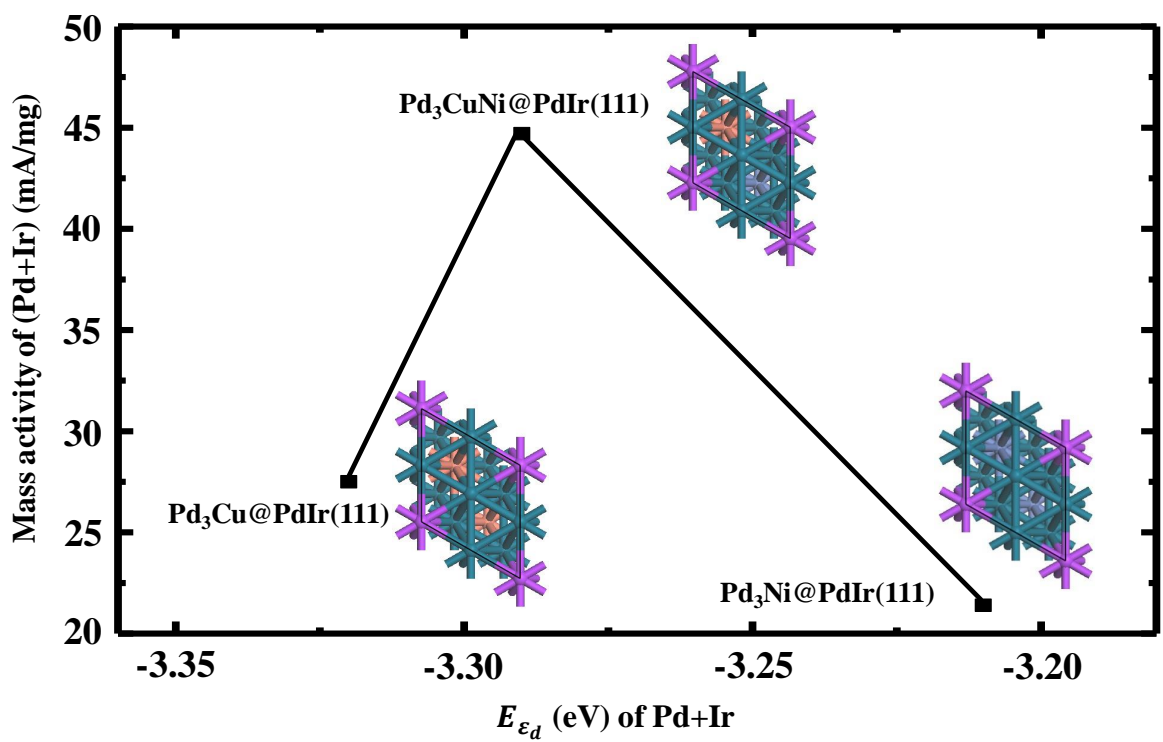

**Figure 6.** Volcano plot of catalytic activity based on the experimentally available mass activities of the catalyst surfaces (both Pd+Ir) and the computationally determined d-band center values for both surface Pd and Ir atoms. Note: the black lines are provided for guidance purposes.

## 4. Conclusions

We constructed simple models of three complicated alloy-core@alloy-shell catalysts, namely Pd$_3$Ni@PdIr, Pd$_3$CuNi@PdIr, and Pd$_3$Cu@PdIr, and investigated the effects of core composition, surface alloying, and compressive strain on dioxygen-adsorption energy. Each core is composed of Pd alloyed with Ni, Cu-Ni, or Cu, and each shell is alloyed with Ir. Experimentally obtained lattice parameters revealed that the Pd$_3$Cu@PdIr catalyst has the smallest lattice constant, followed by Pd$_3$CuNi@PdIr and Pd$_3$Ni@PdIr. These lattice parameter values were assigned to the initial unstrained structures, and -1% to -5% strain was systematically introduced in order to study the effect of compressive strain on the dioxygen-adsorption energy. The calculated dioxygen-adsorption energies for the surface Pd and Ir atoms revealed that the Pd$_3$CuNi@PdIr catalyst has the lowest dioxygen-adsorption-energy magnitude at a given compressive strain. The highest downshift of the d-band center is associated with the shortest bond length on the catalyst surface; consequently, the highest d-band center downshift for the surface Pd atoms is associated with the Pd$_3$Cu@PdIr catalyst, whereas the highest d-band center downshift for the surface Ir atoms was observed for the Pd$_3$CuNi@PdIr catalyst. This difference is reconciled by Bader charge calculations for the surface Pd and Ir atoms, which showed that the lowest amount of charge accumulate on the Pd$_3$CuNi@PdIr catalyst surface, which decreased the magnitude of the dioxygen-adsorption energy. We conclude that through synergism between charge depletion, the d-band center shift, and the surface alloy effect, the Pd$_3$CuNi@PdIr catalyst exhibits the poorest dioxygen adsorption. Finally, a volcano-shaped relationship was produced by plotting the experimentally obtained mass activities for the catalyst surfaces against the theoretically calculated d-band centers of the surface Pd and Ir atoms, in which the Pd$_3$CuNi@PdIr catalyst was at the summit of the volcano. Thus, we correctly reproduced the experimentally observed catalytic-activity trend in which the Pd$_3$CuNi@PdIr catalyst is the most active for the ORR, followed by Pd$_3$Cu@PdIr and Pd$_3$Ni@PdIr. We propose that the catalytic activities of surface Pd atoms can be improved by selecting elements that maximize the electronegativity difference between the surface Pd and the alloyed metal, thereby reducing dioxygen adsorption on the surface alloy. This work sheds light on the importance of the ligand effect, the strain effect, and surface alloying when rationally designing catalysts from first principles by fine-tuning alloy-core@alloy-shell materials.

**Acknowledgements**

This research was supported by the National Research Foundation of Korea (NRF) funded by the Ministry of Science, ICT & Future Planning (Nos. NRF-2016M1A2A2937151 and NRF-2015M1A2A2057129 and NRF-2016M1A2A2937159). This work was supported by the BB21+ Project in 2018.